\documentclass[seceq]{ptptex}
\usepackage{wrapft}
\usepackage{graphicx}

\newcommand{\wtilde}[1]{\widetilde{#1}} 

\newcommand{\lal}{\langle\!\langle}
\newcommand{\rar}{\rangle\!\rangle}
\newcommand{\wb}[1]{\overline{#1}}
\def\beq{\begin{eqnarray}}
\def\eeq{\end{eqnarray}}
\def\bsub{\begin{subequations}}
\def\esub{\end{subequations}}
\def\b{\begin{equation}}

\title{
First Order Quark-Hadron Phase Transition in the NJL-Type Nuclear and 
Quark Model}
\subtitle{The Case of Symmetric Nuclear Matter
}    

\author{
Yasuhiko {\sc Tsue},$^{1}$ 
Jo\~ao da {\sc Provid\^encia}$^{2}$,\\
Constan\c{c}a {\sc Provid\^encia}$^{2}$
and Masatoshi {\sc Yamamura}$^{3}$ 
}

\inst{
$^{1}$Physics Division, Faculty of Science, Kochi University, Kochi 780-8520, 
Japan\\
$^{2}$Departamento de F\'{i}sica, Universidade de Coimbra, 3004-516 Coimbra, 
Portugal\\
$^3$Department of Pure and Applied Physics, 
Faculty of Engineering Science,\\
Kansai University, Suita 564-8680, Japan
}

\recdate{
\today
}

\abst{
The quark-hadron phase transition at finite baryon chemical potential 
is investigated in the extended Nambu-Jona-Lasinio model in which 
the scalar-vector eight-point interaction is included holding the 
chiral symmetry. 
By comparing a pressure of the symmetric nuclear matter with that of
the quark matter, the realized phase at given baryon density is 
determined. 
As a result, the first order quark-hadron phase transition is 
derived where the deconfinement transition occurs after the chiral symmetry restoration 
in this model. 
}

\begin{document}

\maketitle

\section{Introduction}

One of the recent interests in the systems governed by the quantum chromodynamics 
(QCD) may be to understand the phase diagram and/or the phase transitions between 
various phases in the quark-gluon and the hadronic matters. 
As for the chiral phase transition, many works have been done in the various 
effective models of QCD at finite temperature and/or density. 
However, to obtain definite results of the quark-hadron phase transition at finite 
baryon density is still difficult because of the color confinement 
on the side of the hadronic phase. 
Of course, the study based on the lattice QCD is also difficult at finite density at present.

For the symmetric nuclear matter, it is important to describe the property of nuclear 
saturation. 
The Walecka model\cite{Walecka} has succeeded in describing the saturation 
property of symmetric nuclear matter as a relativistic system. 
The success is mainly due to the cancellation of the large scalar and 
vector potential which are derived by the $\sigma$ meson and $\omega$ meson 
exchanges between nucleons, respectively. 
In this model, the nucleon is treated as not a composit but a fundamental particle. 
Although this model has given many successful results for nuclei and nuclear matter, 
this model at first stage has no chiral symmetry which plays an important role in QCD.

The celebrated Nambu-Jona-Lasinio (NJL) model\cite{NJL} gives many important 
results for hadronic world\cite{HK} based on the concenpts of the chiral symmetry 
and the dynamical chiral symmetry breaking. 
This model has been applied to the investigation of the dense quark matter.\cite{BUBALLA}
Also, by using this model, the stability of nuclear matter, as well as quark matter, 
was investigated in which the 
nucleon is constructed from the viewpoint of quark-diquark picture.\cite{BT01} 
On the other hand, it is known that, if the nucleon field is regarded as a fundamental fermion field, 
not composite one, the nuclear saturation property can not be reproduced 
starting from the original NJL Lagrangian. 
However, if the scalar-vector and isoscalar-vector eight-point interactions 
are introduced holding the chiral symmetry in the original NJL model, 
the nuclear saturation property is well reproduced\cite{KBKM87} where the nucleon 
is treated as a fundamental fermion.

In this paper, paying an attention to the chiral symmetry, the NJL-type 
model is adopted in both the nuclear and quark matters. 
As for the nuclear matter, we adopt the extended NJL model with the eight-point 
interactions and regard the nucleon field as a fundamental fermion field with the number of 
color, $N_c$, being one.\cite{MPPM,PPM02} 
Also, we adopt the extended NJL model with $N_c=3$ for the quark matter. 
Thus, the quark-hadron phase transition is investigated by using the same-type model with 
the chiral symmetry for nuclear and quark matters in a unified way, the model 
parameters and the number of color of which are different. 
As the first step to invetigate the quark-hadron phase transition, 
in this paper, we consider the symmetric nuclear matter and free quark phase without quark-pair correlation. 
Namely, we do not take into account of the color superconducting phase 
in the quark phase at finite density and at zero temperature.\cite{cs} 
This phase may exist at finite density, but, in this paper, the investigation is rest 
at future. 
Also, we only consider the symmetric nuclear matter, while it is interesting to 
the study of the neutron matter which leads to the understanding of the physics of 
neutron star.\cite{MP}

This paper, is organized as follows: 
In the next section, we introduce the extended NJL model at finite temperature and baryon 
chemical potential for nuclear and quark matters. 
The mean field approximation in this model is given. 
The derived results are identical to those derived by the minimization 
of the thermodynamical potential density. 
In \S 3, the method to determine the model parameters is presented 
for nuclear matter and quark matter, respectively. 
In \S 4, the numerical results are given. 
In \S 5, the quark-hadron phase transition is described in this model. 
The last section is devoted to a summary and concluding remarks.

\section{Extended NJL model for nuclear and quark matter at finite temperature 
and density}

In this section, the same models for the nuclear and quark matters are 
considered in which parameters are different for nuclear and quark matters. 
One possibe model may be a Nambu-Jona-Lasinio (NJL) 
model\cite{NJL} for nucleons and/or 
quarks. 
The NJL model is useful because it implements in a simple way the important 
chiral symmetry.
As for the nuclear matter, the property of nuclear saturation 
must be realized. 
Then, as one of the possible models, 
we give the NJL-type model in which 
the fermions 
interact with each other through the original NJL-type 
four-point intaraction 
plus four point vector-vector and eight-point scalar-vector and 
isoscalar-vector interactions. 
This model is called an extended NJL model in this paper.

\subsection{Extended NJL model for nuclear and quark matters at finite 
density and temperature}

In this paper, we start with the following Lagrangian 
density\cite{MPPM,KBKM87} for nuclear matter ($i=N$) and  
qurak matter ($i=q$): 
\begin{eqnarray}\label{2-1}
{\cal L}_i&=&{\wb \psi}_i i\gamma^{\mu}\partial_{\mu}\psi_i 
+G_s^i(({\wb \psi}_i{\psi}_i)^2+({\wb \psi}_i i\gamma_5{\mib \tau}\psi_i)^2)
-G_v^i({\wb \psi}_i\gamma^{\mu}\psi_i)({\wb \psi}_i\gamma_{\mu}\psi_i)
\nonumber\\
& &
-G_{sv}^i(({\wb \psi}_i{\psi}_i)^2+({\wb \psi}_i i\gamma_5{\mib \tau}\psi_i)^2)
({\wb \psi}_i\gamma^{\mu}\psi_i)({\wb \psi}_i\gamma_{\mu}\psi_i) \ . 
\end{eqnarray}
Here, $\psi_i$ represents nucleon field ($i=N$) or quark field ($i=q$). 
The first two terms give the 
original NJL Lagrangian density. However, the nuclear matter saturation 
properties are not reproduced if only these two terms are taken into account. 
Thus, we introduce other two terms, following Ref.\citen{KBKM87}. 
Namely, a vector-vector 
repulsive term, whose interaction strength is represented by $G_v^i$, 
and a vector-scalar coupling term, whose interaction strength is represented 
by $G_{sv}^i$, are introduced. 
This model is nonrenormalizable. Thus, it is necessary 
to introduce the cutoff parameter $\Lambda_i$. We adopt a three-momentum 
cutoff scheme.

We apply the mean field approximation to the above Lagrangian density. 
By mean field approximation we mean that we replace bi-linear quantities 
in the fermion fields, 
such as ${\wb \psi}\Gamma \psi$, by 
$\langle {\wb \psi}\Gamma \psi \rangle+({\wb \psi}\Gamma \psi
-\langle {\wb \psi}\Gamma \psi \rangle)$, and keep only linear terms 
in the fluctuation 
$({\wb \psi}\Gamma \psi
-\langle {\wb \psi}\Gamma \psi \rangle)$. 
Here, the symbol $\langle \cdots \rangle$ denotes the expectation 
value or thermal average. 
Only the expectation values of ${\wb \psi}_i i\gamma^\mu\partial_\mu\psi_i$
and ${\wb \psi}_i\psi_i$  
survive at finite temperature and density. 
The mean field Lagrangian density ${\cal L}_i^{MF}$ 
and the mean field Hamiltonian density ${\cal H}_i^{MF}$ 
are easily obtained as 
\begin{eqnarray}\label{2-2}
{\cal L}_i^{MF}&=&{\wb \psi}_i(i\gamma^{\mu}\partial_{\mu}-m_i)\psi
-{\wtilde \mu}_i{\wb \psi}_i\gamma^0\psi_i + C_i \ , \nonumber\\
{\cal H}_i^{MF}&=&-i{\wb \psi}_i{\mib \gamma}\cdot \nabla\psi_i
+m_i{\wb \psi}_i\psi_i
+{\wtilde \mu}_i{\wb \psi}_i\gamma^0\psi_i -C_i \ , \nonumber\\
& &C_i \equiv 
-G_s^i\lal{\wb \psi}_i\psi_i\rar^2+G_v^i\lal{\wb \psi}_i\gamma^0\psi_i\rar^2
+G_{sv}^i\lal{\wb \psi}_i\psi_i\rar^2\lal{\wb \psi}_i\gamma^0\psi_i\rar^2 \ , 
\end{eqnarray}
where we define $m_i$ and ${\wtilde \mu}_i$ as 
\begin{eqnarray}
& &m_i=-2G_s^i\lal{\wb \psi}_i\psi_i\rar
+2G_{sv}^i\lal{\wb \psi}_i\psi_i\rar\lal
{\wb \psi}_i\gamma^0\psi_i\rar^2 \ , 
\label{2-3}\\
& &{\wtilde \mu}_i=2G_v^i\lal{\wb \psi}_i\gamma^0\psi_i\rar
+2G_{sv}^i\lal{\wb \psi}_i\psi_i\rar^2\lal{\wb \psi}_i\gamma^0\psi_i\rar \  
\label{2-4}
\end{eqnarray}
for the nuclear matter ($i=N$) and the quark matter ($i=q$), respectively. 
Here, the symbol $\lal \cdots \rar$ denotes the thermal average which is 
given latter, while in this paper we only treat the zero temperature system. 
Also, the symbol $\langle \cdots \rangle$ will be used 
for the expectation values at zero temperature.

We deal with a finite density system in this paper. 
In order to calculate the 
physical quantities at finite density, 
we introduce the chemical potential $\mu_i$ : 
\begin{eqnarray}\label{2-5}
{\cal H}_i'&=&{\cal H}_i^{MF}-\mu_i\psi_i^\dagger \psi_i\nonumber\\
&=&-i{\wb \psi}_i{\mib \gamma}\cdot \nabla\psi_i+m_i{\wb \psi}_i\psi_i
-{\mu}_i^r{\wb \psi}_i\gamma^0\psi_i -C_i \ , 
\end{eqnarray}
where the effective chemical potential $\mu_i^r$ is defined as
\begin{equation}\label{2-6}
\mu_i^r=\mu_i-{\wtilde \mu}_i
=\mu_i- \left[2G_v^i\lal{\wb \psi}_i\gamma^0\psi_i\rar
+2G_{sv}^i\lal{\wb \psi}_i\psi_i\rar^2\lal{\wb \psi}_i\gamma^0\psi_i
\rar\right] \ . 
\end{equation}
Under this Hamiltonian density, we can calculate physical quantities for 
nuclear matter and/or quark matter.

The expectation values at zero temperature 
can be compactly expressed as 
\begin{eqnarray}\label{2-7}
& &\langle {\wb \psi}_i\psi_i \rangle 
=-\!\!\int\! \frac{d^4p}{i(2\pi)^4}{\rm Tr} (iS_i(p))
=-4N_c^iN_f^i m_i\! \int\! \frac{d^4p}{i(2\pi)^4}\frac{1}{p^2-m_i^2-i\epsilon} 
\ , \nonumber\\
& &\langle {\wb \psi}_i\gamma^0\psi_i \rangle 
=-\!\!\int\! \frac{d^4p}{i(2\pi)^4}{\rm Tr} (\gamma^0iS_i(p))
=-4N_c^iN_f^i \!\int\! \frac{d^4p}{i(2\pi)^4}\frac{p^0}{p^2-m_i^2-i\epsilon} 
\ , \nonumber\\
& &\langle {\wb \psi}_i({\mib \gamma}\cdot{\mib p})\psi_i \rangle 
=-\!\!\int\! \frac{d^4p}{i(2\pi)^4}{\rm Tr} ({\mib \gamma}\cdot{\mib p}iS_i(p))
=-4N_c^iN_f^i m_i\! \int\! \frac{d^4p}{i(2\pi)^4}
\frac{{\mib p}^2}{p^2-m_i^2-i\epsilon} 
\ , \qquad
\end{eqnarray}
where $iS_i(p)=1/(p^{\mu}\gamma_\mu-m_i-i\epsilon)$ 
is the fermion propagator in 
which $p_0$ is replaced into $p_0+\mu_i^r$. 
We introduce $N_f^i$ and $N_c^i$ 
which represent the numbers of flavor and color. 
For nuclear matter, $N_f^N=2$ and $N_c^N=1$ are adopted, and for quark matter 
$N_f^q=2$ and $N_c^q=3$ are adopted. 
Further, we can use imaginary time formalism to deal with the system under 
consideration at finite temperature. 
As is well known, the time component of four-momentum after Wick's 
rotation, $p_4$, can 
be replaced by 
Matsubara's frequency $\omega_n=(2n+1)\pi T$ where $T=1/\beta$ 
is temperature and the integral $\int dp_0/(2\pi)$ is also replaced into 
the Matsubara sum:\cite{Matsubara} 
$p_0\rightarrow i\omega_n$ and $\int dp_0/(2\pi)\rightarrow
iT\sum_{n=-\infty}^{\infty}$. 
As a result, we can calculate 
the above derived quantities at zero temperature with the degeneracy factor 
$\nu_i=2N_f^iN_c^i$ : 
\begin{eqnarray}
& &\lal {\wb \psi}_i\psi_i \rar 
=\nu_i\int \frac{d^3{\mib p}}{(2\pi)^3}\frac{m_i}{\sqrt{{\mib p}^2+m_i^2}}
(n_+^i -n_-^i)\ , 
\label{2-8}\\
& &\lal {\wb \psi}_i\gamma^0\psi_i \rar 
=\nu_i\int \frac{d^3{\mib p}}{(2\pi)^3}
(n_+^i +n_-^i)\ , 
\label{2-9}\\
& &\lal {\wb \psi}_i({\mib \gamma}\cdot{\mib p})\psi_i \rar 
=\nu_i\int \frac{d^3{\mib p}}{(2\pi)^3}
\frac{{\mib p}^2}{\sqrt{{\mib p}^2+m_i^2}}
(n_+^i -n_-^i)\ , 
\label{2-10}
\end{eqnarray}
where $n_{\pm}^i$ is the fermion number distribution functions defined as 
\begin{equation}\label{2-11}
n_{\pm}^i=\left[e^{\beta(\pm\sqrt{{\bf p}^2+m_i^2}-\mu_i^r)}+1\right]^{-1} \ . 
\end{equation}
%
Here, the contribution of the occupied negative energy states should be eliminated from 
the nucleon and/or quark number density itself in Eq.(\ref{2-9}). 
Therefore, we will replace $n_-^i$ by $n_-^i-1$ in Eq.(\ref{2-9}). 
Thus, Eqs.(\ref{2-3}), (\ref{2-6}), (\ref{2-8}) and 
(\ref{2-9}) with (\ref{2-11}) constitute a set of the self-consistent 
equations.

\subsection{Thermodynamical potential density}

Equation (\ref{2-3}) with (\ref{2-8}) and (\ref{2-9}) gives a self-consistent 
equation for $m_i$. 
This is nothing but the so-called gap equation. 
This gap equation (\ref{2-3}) and the fermion number distribution functions 
(\ref{2-11}) presented in the previous subsection are obtained 
by minimizing the thermodynamical potential density $\omega_i$ 
in the 
mean field approximation. Here, $\omega_i$ is defined as 
\begin{equation}\label{2-12}
\omega_i=\lal {\cal H}_i^{MF} \rar -\mu_i\lal {\cal N}_i \rar
-\frac{1}{\beta}
\lal {S_i} \rar , 
\end{equation}
where 
\begin{eqnarray}
& &\lal {\cal H}_i^{MF} \rar=\lal {\wb \psi}_i({\mib \gamma}\cdot{\mib p})
\psi_i \rar 
-G_s^i\lal {\wb \psi}_i\psi_i \rar^2 +G_v^i\lal {\wb \psi}_i\gamma^0\psi_i 
\rar^2
+G_{sv}^i\lal {\wb \psi}_i\psi_i \rar^2\lal {\wb \psi}_i\gamma^0\psi_i \rar^2
\ , \nonumber\\
& &\label{2-13}\\
& &\lal {\cal N}_i\rar=\lal {\wb \psi}_i\gamma^0\psi_i \rar \ , 
\label{2-14}\\
& &\lal S_i\rar=-\nu_i\int\frac{d^3{\mib p}}{(2\pi)^3}\biggl[
n_+^i\ln n_+^i + (1-n_+^i)\ln (1-n_+^i) \nonumber\\
& &\qquad\qquad\qquad\qquad\qquad\qquad\qquad
+n_-^i\ln n_-^i + (1-n_-^i)\ln (1-n_-^i)\biggl] \ 
\label{2-15}
\end{eqnarray}
for nuclear matter ($i=N$) and quark matter ($i=q$), respectively. 
Here the thermodynamical averages 
$\lal {\cal O} \rar={\rm Tr}{\cal O}e^{-\beta
({\cal H}_i^{MF}-\mu_i{\cal N}_i)}/
{\rm Tr}e^{-\beta({\cal H}_i^{MF}-\mu_i{\cal N}_i)}$ 
have the same forms as Eqs.(\ref{2-8})$\sim$ (\ref{2-10}). 
By minimizing $\omega_i$ with respect to $m_i$, $n_+^i$ and $n_-^i$, 
we get the following equations : 
\begin{eqnarray}
& &\frac{\partial \omega_i}{\partial m_i}
=\nu_i\frac{{\mib p}^2}{({\mib p}^2+m_i^2)^{3/2}}(n_+^i -n_-^i)\nonumber\\
& &\qquad\qquad\qquad\times
\left[-m_i-2G_s^i\lal {\wb \psi}_i\psi_i \rar 
+2G_{sv}^i\lal {\wb \psi}_i\gamma^0\psi_i 
\rar^2
\lal {\wb \psi}_i\psi_i \rar\right]=0 \ , 
\label{2-16}\\
& &\frac{\partial \omega_i}{\partial n_+^i}
=\nu_i\biggl[\frac{1}{\sqrt{{\mib p}^2+m_i^2}}\left\{
{\mib p}^2+m_i(-2G_s^i\lal {\wb \psi}_i\psi_i \rar +2G_{sv}^i
\lal {\wb \psi}_i\gamma^0\psi_i \rar^2
\lal {\wb \psi}_i\psi_i \rar)\right\} \nonumber\\
& &\qquad\qquad
-(\mu_i-2G_v^i\lal {\wb \psi}_i\gamma^0\psi_i \rar -2G_{sv}^i
\lal {\wb \psi}_i\gamma^0\psi_i \rar
\lal {\wb \psi}_i\psi_i \rar^2)
+\frac{1}{\beta}\ln\frac{n_+^i}{1-n_+^i}\biggl]=0 \ , 
\nonumber\\
& &\label{2-17}\\
& &\frac{\partial \omega_i}{\partial n_-^i}
=\nu_i\biggl[-\frac{1}{\sqrt{{\mib p}^2+m_i^2}}\left\{
{\mib p}^2+m_i(-2G_s^i\lal {\wb \psi}_i\psi_i \rar +2G_{sv}^i
\lal {\wb \psi}_i\gamma^0\psi_i \rar^2
\lal {\wb \psi}_i\psi_i \rar)\right\} \nonumber\\
& &\qquad\qquad
-(\mu_i-2G_v^i\lal {\wb \psi}_i\gamma^0\psi_i \rar -2G_{sv}^i
\lal {\wb \psi}_i\gamma^0\psi_i \rar
\lal {\wb \psi}_i\psi_i \rar^2)
+\frac{1}{\beta}\ln\frac{n_-^i}{1-n_-^i}\biggl]=0 \ . 
\nonumber\\
& &\label{2-18}
\end{eqnarray}
Thus, we obtain the gap equation (\ref{2-3}) from (\ref{2-16}) again, 
and then, 
the fermion number distribution functions $n_\pm^i$ in (\ref{2-11}) are 
also obtained 
from Eqs.(\ref{2-17}) and (\ref{2-18}) with the effective chemical potential (\ref{2-6}).

\section{Determination of model parameters}

In this section, we give the method to determine the model parameters.

\subsection{For nuclear matter}

In this subsection, we give the prescription to fix the model 
parameters whose numerical values are presented in the next 
section. 
In the model for nuclear matter 
given in the previous section, we have four parameters: 
$G_s^N$, $G_v^N$, $G_{sv}^N$ and the three momentum cutoff $\Lambda_N$. 
We can fix these parameters so as to reproduce the vacuum and saturation 
properties for nucleon and nuclear matter at zero temperature. 
At zero temperature, the fermion number distribution function given in 
Eq.(\ref{2-11}) 
is reduced into 
\begin{equation}\label{3-1}
n_+^N=\theta(\mu_N^r-\sqrt{{\mib p}^2+m_N^2}) \ , \qquad
n_-^N=1 \ , 
\end{equation}
where $\theta$ is the Heaviside step function. 
Here, the contribution of the anti-nucleon for nuclear matter density 
in Eq.(\ref{2-9}) for $i=N$ is 
subtracted as was mentioned in the previous section, namely, 
$n_-^N$ should be replaced to $n_-^N-1$. 
Thus, the nucleon number density (\ref{2-9}) is calculated as 
\begin{eqnarray}\label{3-2}
\rho_N\equiv \langle \psi_N^\dagger \psi_N \rangle
&=&\frac{\nu_N}{6\pi^2}({\mu_N^r}^2-m_N^2)^{3/2}
\nonumber\\
&=&\frac{\nu_N}{6\pi^2}{p_F^N}^3 \ ,  
\end{eqnarray}
where we have introduced the Fermi momentum $p_F^N=\sqrt{{\mu_N^r}^2-m_N^2}$ 
for nuclear matter. 
Then, the gap equation (\ref{2-3}) with (\ref{2-8}) can be expressed as
\begin{eqnarray}\label{3-3}
& &m_N=-2G_s^N\left(1-\frac{G_{sv}^N}{G_s^N}\rho_N^2\right) 
\langle {\wb \psi}_N\psi_N \rangle \ , \nonumber\\
& &\langle {\wb \psi}_N\psi_N \rangle =-\frac{\nu_N m_N}{2\pi^2}
\int_{p_F^N}^{\Lambda_N} d|{\mib p}|\frac{{\mib p}^2}{\sqrt{{\mib p}^2+m_N^2}} 
\ .
\end{eqnarray}
It is seen from (\ref{3-3}) that the scalar-vector coupling term 
in the Lagrangian density (\ref{2-1}) gives the density dependent 
coupling $G_s^N(\rho_N)\ (=G_s^N(1-G_{sv}^N/G_s^N\cdot \rho_N^2))$ in the  
gap equation of the original NJL model Lagrangian with 
$G_{sv}^N=G_{v}=0$. 
Also, $\langle{\wb \psi}_N({\mib \gamma}\cdot{\mib p})\psi_N\rangle$ is 
calculated as 
\begin{equation}\label{3-4}
\langle{\wb \psi}_N({\mib \gamma}\cdot{\mib p})\psi_N\rangle
=-\frac{\nu_N}{2\pi^2}\int_{p_F^N}^{\Lambda_N}d|{\mib p}|\frac{|{\mib p}|^4}{
\sqrt{{\mib p}^2+m_N^2}} \ .
\end{equation}
Further, the energy density per single nucleon at finite baryon density 
and zero temperature is easily evaluated as 
\begin{equation}\label{3-5}
W_N(\rho_N)=\frac{\langle {\cal H}_N^{MF}\rangle (\rho_N)-
\langle {\cal H}_N^{MF}\rangle (\rho_N=0)}{\rho_N}-m_N(\rho_N=0) \ ,
\end{equation}
where 
\begin{equation}\label{3-6}
\langle {\cal H}_N^{MF}\rangle (\rho_N)
=\langle {\wb \psi}_N({\mib \gamma}\cdot{\mib p})\psi_N\rangle
-G_s^N(1-G_{sv}^N/G_s^N\cdot \rho_N^2)\langle {\wb \psi}_N\psi_N\rangle^2
+G_v^N\rho_N^2 \ .
\end{equation}
Here, we have denoted the density dependence explicitly.

From (\ref{3-3}), we fix the nucleon mass at zero and normal nuclear 
density as 
$m_N(\rho_N=0)=939$ MeV and $m_N(\rho_N^0)=0.6 m_N(\rho_N=0)$, respectively. 
Here, $\rho_N^0=0.17$ fm$^{-3}$ is a normal nuclear density 
and $m_N(\rho_N^0)$ is given as a typical effective mass value at the normal 
nuclear density. 
The reason why we take the effective nucleon mass 
$m_N(\rho_N^0)=0.6 m_N(\rho_N=0)$ is that the quark mass in the 
original NJL model with $G_{sv}^q=G_v^q=0$ can 
be numerically calculated as 187 MeV at $\rho_N^0$. 
Thus, the nucleon mass at $\rho_N^0$ is approximately evaluated as 
$561=3\times 187$ MeV which is about $0.6 m_N$ ($\approx 563$ MeV). 
From (\ref{3-5}), we fix the parameters so as to reproduce the 
nuclear matter saturation properties, namely, $W_N$ should give 
the minimum value $-15$ MeV at normal nuclear density $\rho_N^0$. 
Then, our model-parameters are determined by these four conditions:
nucleon mass at zero and normal nuclear density and the saturation properties. 
As will be shown in \S\S 4.1., the nucleon mass at normal nuclear density gives 
an influence to the imcompressibility of nuclear matter at normal 
nuclear density. 
Therefore, it is possible that the imcompressibility of nuclear matter at 
normal nuclear density is adopted as an input parameter instead of the nucleon 
mass at normal nuclear density. 
However, in this paper, we fix the value of the nucleon mass at the normal 
nuclear density derived by the simple quark model which is implemented by the 
quark NJL model.

\subsection{For quark matter}

For quark matter, we use same model Lagrangian density (\ref{2-1}) as 
that used in the nuclear matter with different model parameters. 
For quark matter, 
the parameters $G_v^q$ and $G_{sv}^q$ are usually set equal to 0, because, in nuclear 
matter, the corresponding parameters were 
introduced so as to reproduce the nuclear matter 
saturation properties. 
Thus, we put $G_v^q=0$. 
However, we retain $G_{sv}^q$ which appears in the gap equation (\ref{2-3}) 
at finite density. 
Then, the scalar-vector and isoscalar-vector interaction term with 
$G_{sv}^q$ gives an influence to the chiral phase transition at finite 
density. 
Namely, $G_{sv}^q$ may be used to tune the chiral phase transition and the 
slope of the equation of state at high densities, while 
$G_{sv}^q$ is treated as a free parameter in this paper.

Then, there are three model parameters, namely, $G_s^q$, 
the three momentum cutoff $\Lambda_q$ and $G_{sv}^q$. 
As for the $G_s^q$ and $\Lambda_q$, these two parameters are 
determined by giving the dynamical quark mass $m_q$, which is obtained by 
solving the gap equation (\ref{2-3}) in the vacuum, and 
the pion decay constant $f_\pi$: 
\begin{eqnarray}\label{3-7}
& &m_q=-2G_s^q\langle {\wb \psi}_q \psi_q \rangle 
=\frac{2G_s^q N_cN_f m_q}{\pi^2}\int_0^{\Lambda_q}d|{\mib p}|
\frac{{\mib p}^2}{\sqrt{{\mib p}^2+m_q^2}} \ , \nonumber\\
& &f_\pi^2=\frac{N_c m_q^2}{2\pi^2}\int_0^{\Lambda_q}d|{\mib p}|
\frac{{\mib p}^2}{({\mib p}^2+m_q^2)^{3/2}} \ .
\end{eqnarray}
Then, $G_s^q$ and $\Lambda_q$ are taken so as to reproduce 
$m_q=313$ MeV and $f_\pi=93$ MeV in the vacuum.

Here, $G_{sv}^q$ is taken as a free parameter in this paper. 
The change of $G_{sv}^q$ leads to the change of the chiral phase transition 
point as will be given later.

The physical quantities at finite density and temperature 
can be calculated similar to the case of the nuclear matter. 
Then, $m_N$, $\mu_N^r$ and $\nu_N$ should read 
$m_q$, $\mu_q^r$ and $\nu_q=2N_f^q N_c^q$ with $N_f^q=2$ and $N_c^q=3$, 
respectively, 
and $G_v^q=0$. 
Here, $\mu_q$ is a quark chemical potential. 
Of course, in the previous expressions, $n_-^q$ should be  
replaced to $n_-^q -1$ in the expression of the quark number density.

\section{Numerical Results}

\subsection{Nuclear Matter}

For nuclear matter, we take the parameters so as to reproduce the 
nucleon mass at normal nuclear density as $m_N(\rho_N^0)
=0.6 m_N(\rho_N=0)$. Other conditions are fixed as follows: 
The nucleon mass in vacuum, $m_N=939$ MeV, and 
the saturation properties of infinite nuclear matter, 
$W_N=-15$ MeV at the normal nuclear density $\rho_N^0=0.17$ fm$^{-3}$.
Then, 
the parameter set is summarized in Table I.  
\begin{table}[b]
\caption{The values of model parameters for nuclear matter.}
\label{table:1}
\begin{center}
\begin{tabular}{c|c} \hline \hline
$\qquad\Lambda_N\qquad$ & $\qquad$ 377.8 [MeV] $\qquad$\\
$G_s^N\Lambda_N^2$ & 19.2596 \\
$G_{sv}^N\Lambda_N^8$ & $-1069.89$ \\
$G_v^N\Lambda_N^2$ & 17.9824 \\
\hline
\end{tabular}
\end{center}
\end{table}
\begin{table}[b]
\caption{The values of model parameters for quark matter.}
\label{table:2}
\begin{center}
\begin{tabular}{c|c} \hline \hline
$\qquad\Lambda_q\qquad$ & $\qquad$ 653.961 [MeV] $\qquad$\\
$G_s^q\Lambda_q^2$ & 2.13922 \\
\hline
\end{tabular}
\end{center}
\end{table}

\begin{figure}[t]
\begin{center}
\includegraphics[height=5.3cm]{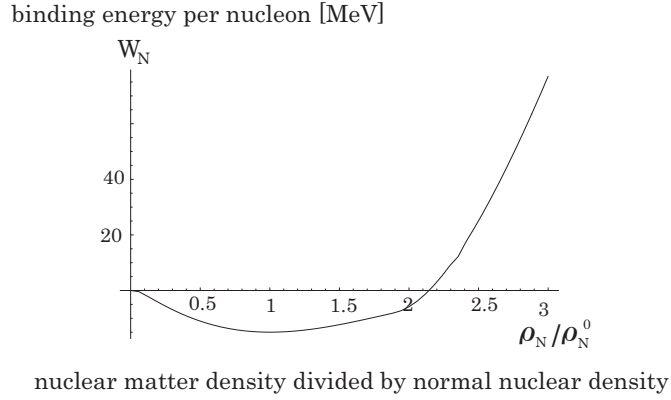}
\caption{Nuclear matter saturation properties with 
$m_N(\rho_N^0)=0.6 m_N(\rho_N=0)$:
Parameters are summarized in Table I. 
Then, the imcompressibility is $K=260$ [MeV].
}
\label{fig:4-1}
\end{center}
\end{figure}

In these model parameters, the incompressibility of the nuclear matter 
at the normal nuclear density is numerically evaluated as 
\beq\label{4-1}
K=9{\rho_N^0}^2\frac{d^2 W_N(\rho_N)}{d\rho_N^2}\biggl|_{\rho_N=\rho_N^0}
\approx 260\ {\rm [MeV]} \ .
\eeq
Thus, a rather reasonable value is obtained. 
In Fig.\ref{fig:4-1}, the energy density per single nucleon in Eq.(\ref{3-5}) 
is depicted as a function of the nuclear matter density divided by 
the normal nuclear matter density $\rho_N^0$, namely, $\rho_N/\rho_N^0$.

However, 
the momentum cutoff $\Lambda_N$ is rather small. 
This leads to the question 
of the applicability of this model. 
The momentum of nucleon should be smaller than the value of three-momentum 
cutoff. We thus do not take into account of 
the scalar excitation of nucleon and antinucleon 
whose mass is the twice of nucleon mass.

If we take the nucleon mass value at the normal nuclear density as 
$m_N(\rho_N^0)=0.75 m_N(\rho_N=0)$, then the parameters are obtained 
as 
$\Lambda_N=441.2$ MeV, $G_s^N\Lambda_N^2=16.7483$, 
$G_{sv}^N\Lambda_N^8=-1958.12$ and $G_v^N\Lambda_N^2=16.9872$.
Then, the imcompressibility is obtained as $K=286$ MeV. 
Thus, the equation of state becomes stiff at normal nuclear density. 

\subsection{Quark Matter}

For quark matter, 
we set $G_v^q=0$ in Eq.(\ref{2-1}) as was already mentioned in the preceeding 
section. 
As for the model parameters, the coupling constant $G_s^q$ and the three 
momentum cutoff $\Lambda_q$ are determined from the dynamical quark mass 
and the pion decay constant in the vacuum in Eq.(\ref{3-7}). 
The parameters are summarized in Table II.

\subsubsection{$G_{sv}^q=0$}

\begin{figure}[t]
\begin{center}
\includegraphics[height=5cm]{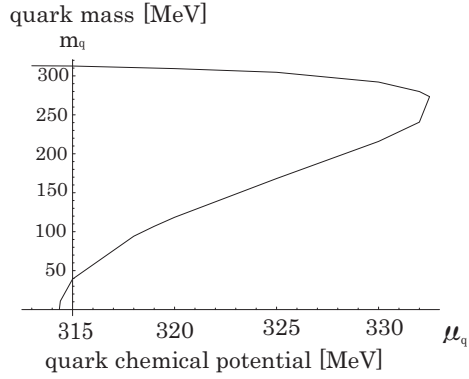}
\caption{The solutions of the gap equations for quark mass are depicted as 
a function of the quark chemical potential $\mu_q$ with $G_{sv}^q=0$ for 
the original NJL model. 
The vertical axis represents the dynamical quark mass $m_q$ and 
the horizontal axis represents the quark chemical potential $\mu_q$. 
}
\label{fig:1-1}
\end{center}
\end{figure}

\begin{figure}[t]
\begin{center}
\includegraphics[height=5cm]{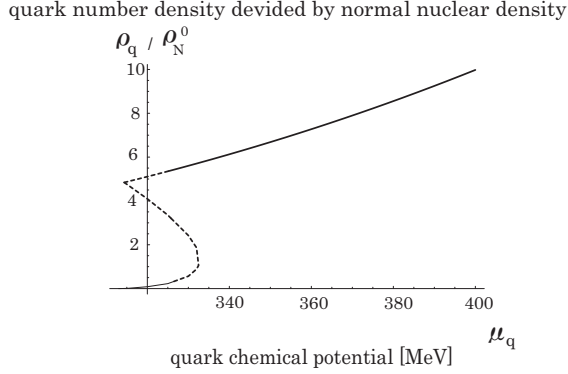}
\caption{The quark number density versus quark chemical potential 
is depicted at $T=0$. The dashed curve is not realized physically. 
}
\label{fig:1-2}
\end{center}
\end{figure}

First, we use the original NJL model Lagrangian density without 
the vector-scalar ($G_{sv}^q$) interactions, namely, we 
put $G_{sv}^q=0$. 
%
%
\begin{figure}[t]
\begin{center}
\includegraphics[height=5cm]{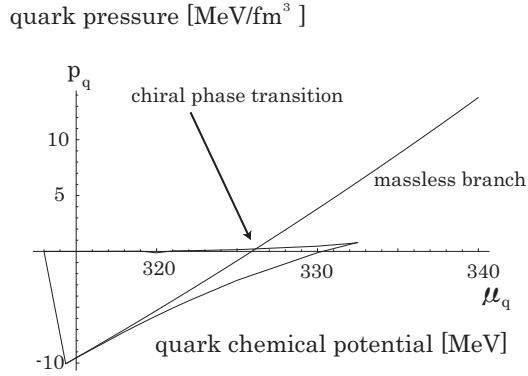}
\caption{The pressure versus chemical potential is depicted at $T=0$. 
The chiral phase transition occurs at $\mu_q \approx 326$ MeV.
}
\label{fig:1-3}
\end{center}
\end{figure}
%
%
We solve the gap equation in Eq.(\ref{2-3}) with $G_{sv}^q=0$ 
under a finite quark chemical potential $\mu_q$ at zero temperature. 
Then, the dynamical quark mass $m_q$ is obtained 
at finite quark chemical potential. 
In Fig.{\ref{fig:1-1}}, the dynamical quark mass is depicted as a function 
of the quark chemical potential. 
For $\mu_q=313$ MeV, the vacuum value for the dynamical quark mass 
is obtained. 
From Eq.(\ref{2-9}) at zero temperature, we obtain the relation 
between the quark number density $\rho_q$ and the quark chemical potential 
$\mu_q$. 
In Fig.{\ref{fig:1-2}}, the quark number density is depicted as a function 
of the quark chemical potential. 
The vertical axis represents the quark number density divided by the normal 
nuclear density, namely, $\rho_q/\rho_N^0$. 
The dashed curves represent the unphysical region. 
Namely, the gap equation has multiple solutions in a certain region 
as is seen from Fig.{\ref{fig:1-1}}. 
Then, we must determine which solution is realized physically. 
For this purpose, 
the pressure is calculated from the thermodynamical potential density 
$\omega$ in Eq.(\ref{2-12}) by the same subtraction method as the energy 
density per single nucleon in Eq.(\ref{3-5}):
\beq\label{4-2}
& &-p_i(T, \mu_i)=
(\lal{\cal H}_i^{MF}\rar(T, \mu_i)-\lal{\cal H}_{i}^{MF}\rar(T=0,\mu_i=m_i(T=0)))
\nonumber\\
& &\qquad\qquad\qquad
-\mu_i\lal{\cal N}_i\rar-\frac{1}{\beta}\lal{\cal S}_i\rar  \ ,
\eeq
where $i=q$ for the quark matter. 
The physical solution corresponds to the one which gives the largest pressure. 
In Fig.{\ref{fig:1-3}}, the pressure $p_q$ is depicted as a function of 
the quark chemical potential $\mu_q$. 
From $\mu_q=313$ MeV to about $326$ MeV, 
the low density solution is realized 
seen from Fig.{\ref{fig:1-2}}. 
However, above $\mu_q\approx 326$ MeV, the massless solution is physically 
realized. 
From Fig.{\ref{fig:1-3}}, it is seen that 
the chiral phase transition occurs at $\mu_q \approx 326$ MeV. 
In this case, in the region 
from $\rho_q \sim 0.28 \rho_N^0$ to $\rho_q \sim 5.41 \rho_N^0$ 
(from $\rho_B \sim 0.09 \rho_N^0$ to $\rho_B \sim 1.80 \rho_N^0$ 
where $\rho_B$ represents the baryon number density) 
as is seen in Fig.{\ref{fig:1-2}}, 
the first order chiral phase transition is realized and 
quark phases coexistence occurs. 
It seems that this density is rather small compared with the 
quark-hadron phase transition as is mentioned below.

%
%

\subsubsection{$G_{sv}^q\Lambda_q^8=-68.4$}

\begin{figure}[t]
\begin{center}
\includegraphics[height=4.7cm]{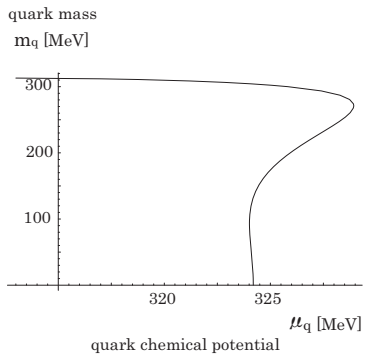}
\caption{The solutions of the gap equations for quark mass are depicted as 
a function of the quark chemical potential $\mu_q$ 
at $T=0$ with $G_{sv}^q\Lambda_q^8=-68.4$. 
}
\label{fig:2-1}
\end{center}
\end{figure}
\begin{figure}[t]
\begin{center}
\includegraphics[height=4.7cm]{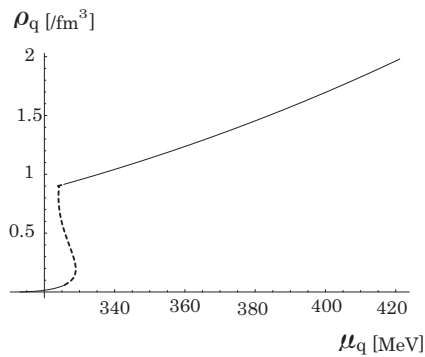}
\caption{The quark number density $\rho_q$ versus quark 
chemical potential $\mu_q$ is depicted at $T=0$ with $G_{sv}^q
\Lambda^8=-68.4$. The dashed curve is not realized physically. 
}
\label{fig:2-2}
\end{center}
\end{figure}

In the original NJL model with $G_{sv}^q=0$, 
it seems that the chiral phase transition 
occurs at rather small density. 
One of possible reasons why the chiral phase transition occurs at rather 
small density may be that the strength of the attractive interaction between 
quark and antiquark is weak. 
So, the quark condensate is melt at rather small density. 
Thus,  there exists one possibility that the attractive interaction 
becomes strong by introducing the scalar-vector attractive 
interaction $G_{sv}^q$ 
for the original quark NJL model.

However, there is no criterion to determine the value of $G_{sv}^q$ in 
this stage. 
Thus, as is the first attempt, we regard $G_{sv}^q$ as a free parameter. 
First, we give the value $G_{sv}^q\Lambda_q^8=-68.4$, in which 
$m_q(\rho_q/3=\rho_N^0)=0.625 m_q(\rho_q=0)$.

\begin{figure}[t]
\begin{center}
\includegraphics[height=4.7cm]{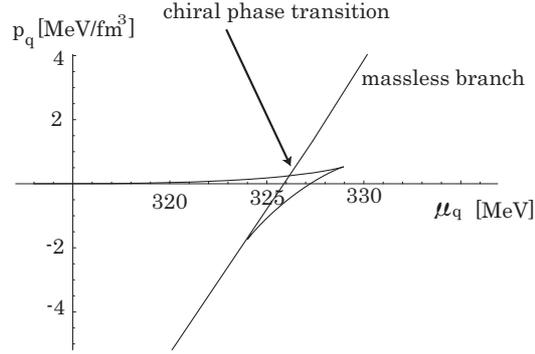}
\caption{The pressure versus quark chemical potential is depicted and 
the chiral phase transition is shown at $T=0$ with $G_{sv}^q
\Lambda^8=-68.4$. 
}
\label{fig:2-3}
\end{center}
\end{figure}

In Fig.{\ref{fig:2-1}}, the dynamical quark mass is depicted as a function 
of the quark chemical potential. 
For $\mu_q=313$ MeV, the vacuum value for the dynamical quark mass 
is also obtained. 
In Fig.{\ref{fig:2-2}}, the quark number density $\rho_q$ itself 
is depicted as a function 
of the quark chemical potential $\mu_q$. 
The dashed curves represent the unphysical region, 
as is similar to Fig.{\ref{fig:1-2}}.  
Namely, the gap equation has multiple solutions in a certain region. 
Then, we must determine which solution is realized physically by comparing 
with each pressure. 
In Fig.{\ref{fig:2-3}}, the pressure for each branch is depicted as a function 
of the quark chemical potential $\mu_q$. 

From Fig.{\ref{fig:2-3}}, it is seen that 
the chiral phase transition occurs at $\mu_q \approx 326$ MeV. 
In this case, in the region 
from $\rho_q \sim 0.38 \rho_0$ to $\rho_q \sim 5.41 \rho_0$ 
(from $\rho_B \sim 0.12 \rho_0$ to $\rho_B \sim 1.80 \rho_0$), 
the chiral phase transition is realized and quark phases coexistence occurs. 

\subsubsection{$G_{sv}^q\Lambda_q^8=-225$}

\begin{figure}[b]
\begin{center}
\includegraphics[height=4.7cm]{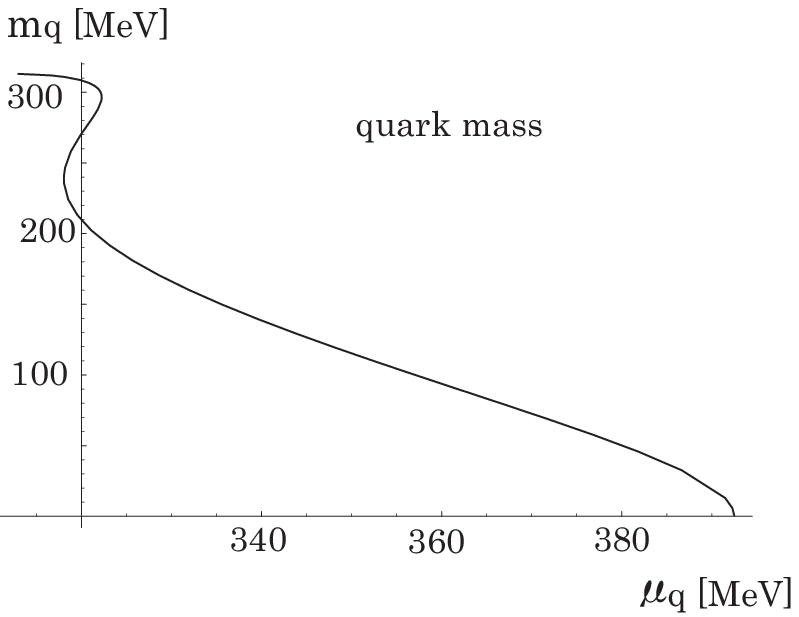}
\caption{The solutions of the gap equations for quark mass are depicted as 
a function of the quark chemical potential $\mu_q$ 
at $T=0$ with $G_{sv}^q\Lambda^8=-225$. 
}
\label{fig:3-1}
\end{center}
\end{figure}

\begin{figure}[t]
\begin{center}
\includegraphics[height=4.7cm]{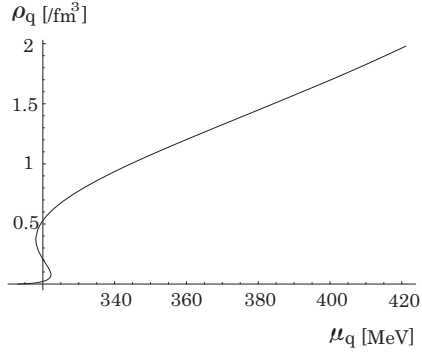}
\caption{The quark number density $\rho_q$ versus quark 
chemical potential $\mu_q$ is depicted at $T=0$ with $G_{sv}^q
\Lambda^8=-225$. 
}
\label{fig:3-2}
\end{center}
\end{figure}

\begin{figure}[t]
\begin{center}
\includegraphics[height=4.7cm]{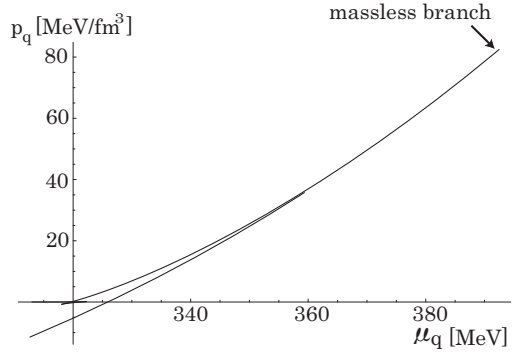}
\caption{The pressure versus quark chemical potential is depicted and 
the chiral phase transition is shown at $T=0$ with $G_{sv}^q\Lambda^8=-225$.
}
\label{fig:3-3}
\end{center}
\end{figure}

Finally, we give the value $G_{sv}^q\Lambda_q^8=-225$ which leads to 
the dynamical quark mass at normal nuclear density as 
$m_q(\rho_q/3=\rho_N^0)=0.681 m_q(0)$. 

In Fig.{\ref{fig:3-1}}, the dynamical quark mass is depicted as a function 
of the quark chemical potential. 
In Fig.{\ref{fig:3-2}}, the quark number density $\rho_q$ 
is depicted as a function 
of the quark chemical potential $\mu_q$. 
In Fig.{\ref{fig:3-3}}, the pressure of each branch is depicted as a function 
of the quark chemical potential. 
From Fig.{\ref{fig:3-3}}, it is seen that 
the phase with massless quark is realized in the region of 
$\mu_q > 392.5$ MeV. 
In this case, at $\rho_q \sim 9.43 \rho_N^0$ 
($\rho_B \sim 3.14 \rho_N^0$), 
the chiral phase transition may be realized. 

\section{Quark-hadron phase transition}

The main purpose of this paper is to investigate the quark-hadron 
phase transition in the extended NJL model developed in this paper. 
In this section, we investigate the realized phase by comparing the 
pressures of the nuclear and quark matters at zero temperature and the 
finite baryon chemical potential. 
It is shown that the first order quark-hadron phase transition 
is realized at zero temperature and the finite baryon chemical potential. 
In order to investigate the phase transition, 
the condition of the chemical equilibrium is demanded for 
the chemical potential $\mu_N$ and $\mu_q$. 
Here, the chemical potential per one baryon at the same 
baryon density should be considered for the chemical equilibrium as follows:
\beq\label{5-1}
\mu_N(T)=3\mu_q(T) \ . 
\eeq
Also, the same pressures between the nuclear matter $p_N$ 
and the quark matters $p_q$ 
are necessary as 
\begin{equation}\label{5-2}
p_N(T,\mu_N)=p_q(T,3\mu_q) \ .
\end{equation}
The pressure has already been defined in Eq.(\ref{4-2}).

The crossing point of each pressure for nuclear matter and the quark matter 
presents the coexistence point of the nuclear and the quark phases and 
the phase with larger value of the pressure 
is realized except for the crossing point.

\begin{figure}[t]
\begin{center}
\includegraphics[height=4.7cm]{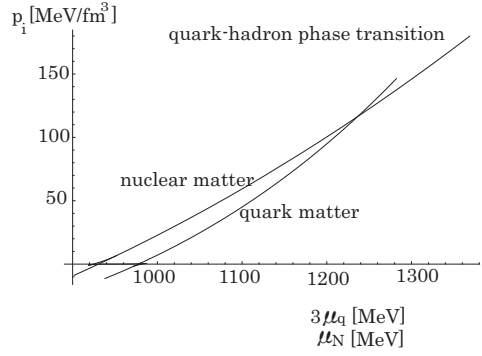}
\caption{The quark-hadron phase transition is shown 
at $T=0$ with $G_{sv}^q\Lambda^8=-68.4$. 
}
\label{fig:5-1}
\end{center}
\end{figure}
\begin{figure}[t]
\begin{center}
\includegraphics[height=5.5cm]{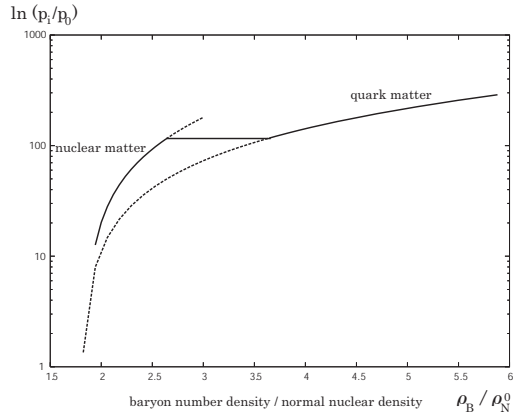}
\caption{The quark-hadron phase transition is shown 
at $T=0$ with $G_{sv}^q\Lambda^8=-68.4$. 
The vertical axis and the horizontal axis represent the 
pressure $\ln (p_i/p_0)$ and the baryon number density divided by the normal 
nuclear density $\rho_B/\rho_N^0$, respectively. Here, we use $p_0=1.0$ MeV/fm$^3$.
}
\label{fig:5-2}
\end{center}
\end{figure}

In Fig.\ref{fig:5-1}, the pressures for the nuclear matter and the 
quark matter are depicted as functions of the nuclear chemical potential 
$\mu_N$ and the quark chemical potential multiplied by 3, $3\mu_q$, 
respectively, in the case $G_{sv}^q\Lambda_q^8=-68.4$. 
In this figure, it is shown that the quark-hadron phase transition occurs 
at a critical chemical potential value $\mu_c$ which is 
about $\mu_c (= \mu_N=3\mu_q) \approx 1236$ MeV. 
From this figure, 
the nuclear or hadron phase is realized in the small chemical potential 
region $\mu_N < \mu_c$. 
However, in the large chemical potential region $3\mu_q > \mu_c$, 
the quark phase is realized. 
In terms of the baryon number density $\rho_B$ 
by using the relation of the density and 
chemical potential shown in Fig.{\ref{fig:2-2}}, 
from $\rho_q \sim 10.9 \rho_0\ (\rho_B=\rho_q/3 \sim 3.63 \rho_0)$ 
to $\rho_N (=\rho_B) \sim 2.65 \rho_0$, 
the nuclear and quark phases coexist and 
the first order quark-hadron phase transition is realized. 
This situation is depicted in Fig.{\ref{fig:5-2}}. 

For the other parameter values for $G_{sv}^q$, the same figure is obtained. 
The reason is as follows: 
For these model parameters used in this paper, that is, 
$G_{sv}^q\Lambda_q^8=0$, $-68.4$ and $-225$, the quark-hadron phase transition point 
is in the chiral symmetric phase, that is, $m_q=0$ and 
$\langle {\overline \psi}_q\psi_q\rangle=0$. 
Namely, the quark-hadron phase transition from hadron phase to quark phase 
occurs after the chiral phase transition from chiral broken phase 
to chiral symmetric phase in these model parameters. 
This behavior for the phase transitions is also seen in Ref.\citen{12}. 
Then, the pressure of quark matter $p_q$ in Eq.(\ref{4-2}) with $T=0$ does not 
depend on $G_{sv}^q$ because 
$\langle {\cal H}_q^{MF} \rangle$ in Eq.(\ref{2-13}) and $\mu_q^r$ in Eq.(\ref{2-6}) 
does not depend on $G_{sv}^q$ due to $\langle {\overline \psi}_q\psi_q\rangle=0$. 
Thus, the behavior seen in Fig.{\ref{fig:5-2}} is not changed.



\section{Summary and concluding remarks}

In this paper, the quark-hadron phase transition at finite baryon 
chemical potential has been described in the extended NJL model, 
in which both the nucleon in the nuclear matter and the quark in the 
quark matter were treated as fundamental fermions with a number 
of color, $N_c$, being one for nucleon and being three 
for quark. 
In this paper, as the first attempt of the investigation for the 
quark-hadron phase transition in the extended NJL model, 
we only dealt with the symmetric nuclear matter and the quark matter without 
quark-quark correlation leading to the pairing instability. 

As for the symmetric nuclear matter, the saturation property has been well reproduced in this model. 
As for the quark matter, there is one free parameter, that is, the 
coupling strength of the scalar-vector eight-point interaction, $G_{sv}^q$. 
This model parameter controlls the chiral phase transition point and/or 
the strength of the partial restoration of the chiral symmetry in the 
nuclear medium. 
In this paper, we did not fix its value by using other 
physical quantity. 
This is a future problem.

We have calculated the pressures of nuclear matter and quark matter, respectively. 
Then, by comparing the pressure of nuclear matter with that of quark matter, 
the realized phase is determined. 
As a result, the first order quark-hadron phase transition is 
obtained at finite density and zero temperature. 
It should be noted here that, for the adopted parameter $G_{sv}^q$ used in this 
paper, the deconfinement phase transition occurs after the chiral symmetry restoration 
in the nuclear matter. 
Thus, the phase transitions in the quark phases may be  hidden 
because they occur before quark deconfinement.

It is interesting to investigate the quark-hadron phase transition 
at finite temperature and baryon chemical potential because 
the phase diagram in the QCD world is not understood completely at present. 
It is also interesting to study the phase transition between the neutron matter 
and quark matter at finite density, which has important implications to the 
physics of neutron star. 
Of course, on the side of quark phase, the color superconducting phase 
should be taken into account. 
These are future problems in this model calculation.

\section*{Acknowledgement} 

One of the authors (Y.T.) would like to express his sincere thanks to 
Professor\break
J. da Provid\^encia and Professor C. Provid\^encia, two of co-authors of this paper, 
for their warm hospitality during his visit to Coimbra in spring of 2009. 
One of the authors (Y.T.) 
is partially supported by the Grants-in-Aid of the Scientific Research 
No. 18540278 from the Ministry of Education, Culture, Sports, Science and 
Technology in Japan.



\end{document}